# Acoustic amplifying diode using non-reciprocal Willis coupling

Xinhua Wen[1], Heung Kit Yip[1], Choonlae Cho[2], Jensen Li[1]*, Namkyoo Park[2]*

[1]*Department of Physics, The Hong Kong University of Science and Technology, Clear Water Bay, Kowloon, Hong Kong, China.*

[2]*Photonic Systems Laboratory, Department of Electrical and Computer Engineering, Seoul National University, Seoul 08826, South Korea*

We propose a concept called acoustic amplifying diode in combining both signal isolation and amplification in a single device. The signal is exponentially amplified in one direction with no reflection and is completely absorbed in another. In this case, the reflection is eliminated from the device in both directions due to impedance matching, preventing backscattering to the signal source. Here, we experimentally demonstrate the amplifying diode using an active metamaterial with non-reciprocal Willis coupling. We also discuss the situation with the presence of both reciprocal and non-reciprocal Willis couplings for more flexibility in implementation. The concept of acoustic amplifying diode will enable applications in sound isolation, sensing and communication, in which non-reciprocity can play an important role.

* jensenli@ust.hk, nkpark@snu.ac.kr

While bianisotropy is a well-known concept in electromagnetism to obtain asymmetric absorption, topological phenomena and power-efficient wavefront shaping [1-4], its counterpart in acoustics being called Willis coupling has only been recently realized through the notion of acoustic metamaterials [5-7]. Consequently, asymmetric or unidirectional zero reflections, non-Hermitian exceptional points can be revealed using acoustic Willis coupling [7-9]. By using asymmetric metamaterial structures with strong local resonances, a significant size of Willis coupling can be achieved [5,6]. However, a maximum bound for such coupling exists for implementations constrained with passivity and reciprocity [10-13]. Breaking the time-reversal symmetry, with moving fluid [14,15] or active (non-Hermitian) components [16-19], can be a solution to generate non-reciprocal Willis coupling and to overcome these constraints. Specifically, programmable metamaterial approach can implement tailor-made impulse response of metamaterials by connecting microphones and speakers through a digital feedback circuit at each meta-atom [20-23]. These meta-atoms mimic how physically resonating metamaterials respond but now with programmable control on the constitutive parameters. By controlling the response of both monopolar and dipolar scattering, Willis coupling beyond the passivity bound can be achieved [19]. The two Willis coupling terms in the constitutive matrix can be controlled to become either symmetric, antisymmetric or arbitrary, making non-reciprocal Willis coupling assessable. Furthermore, the direct and very flexible specifications of the constitutive parameters using these programmable metamaterials, while leaving out the implementation details, may allow us to have an "inverse design" approach starting from effective medium description to formulate one-way acoustic devices, including non-reciprocal lenses and acoustic diodes [14]. In fact, acoustic diodes were demonstrated using nonlinear sonic crystals with frequency conversion [24-26] and using diffraction structures with transmission mode conversion [27]. However, confining to a single channel, no matter regarding frequency or spatial mode, and with significant transmission and zero reflection will be beneficial in order to have an ideal acoustic isolation. Such a single-channel demonstration is still not reachable.

Here, we propose a concept called amplifying diode in combining acoustic isolation and amplification together. The transmitted signal in one propagation direction is exponentially amplified while the transmitted signal in the reverse direction is exponentially suppressed. At the same time, there will be zero reflection from both directions, i.e. an amplifier in one direction and an absorber in another. We design and realize an active and non-reciprocal Willis metamaterial by first investigating the required constitutive parameters for such amplifying diode action and a realization scheme through programmable metamaterials. We will also discuss realization schemes with both the reciprocal and non-reciprocal Willis coupling present. The integration of amplifying action into an ideal diode will be potentially useful to enrich the applications of sound isolation and further enable ultrasensitive non-reciprocal sensing and non-reciprocal communication.

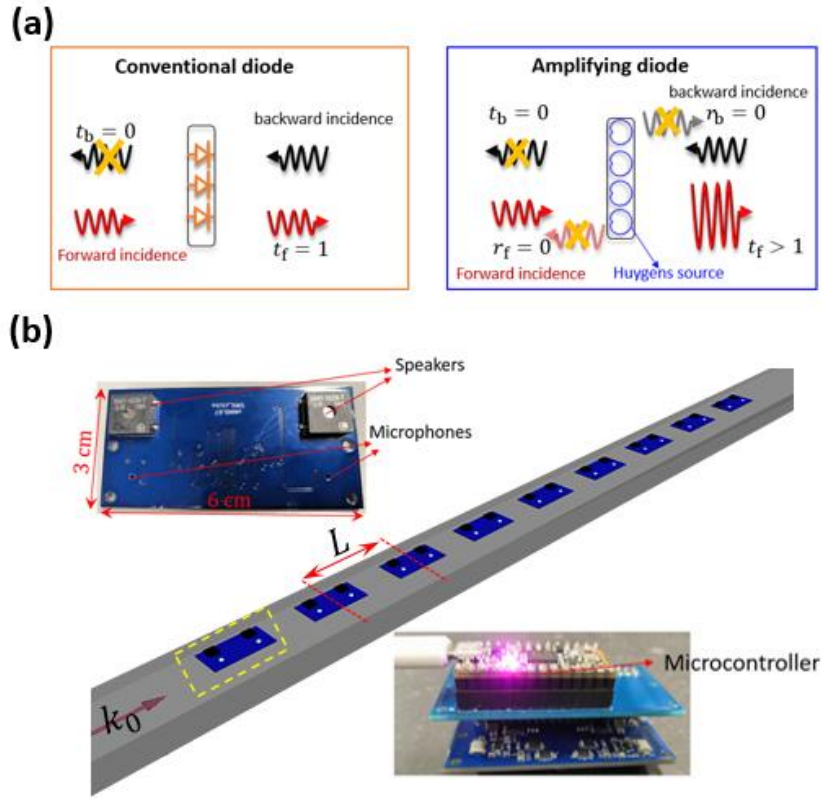

**Figure 1** (a) Schematics of the conventional diode (left panel) and the amplifying diode (right panel). The conventional diode offers a unit transmission amplitude in the forward direction (red color), but a completely suppressed transmitted signal (black color) in the reverse direction. While the amplifying diode offers an amplified forward transmitted signal ($t_f > 1$) but a zero transmission amplitude in the backward direction, exhibiting a significant isolation effect together with amplfiication. Meanwhile, the reflected signals in both the forward and backward directions are completely suppressed. (b) Schematic of the amplifying diode consisting of 9 meta-atoms in a 1D waveguide. The upper left inset shows the photograph of the inverse side of a homemade PCB sample integrated with two speakers and two microphones, and the lower right inset shows an assembled meta-atom connected to the microcontroller.

We begin with the conceptual relationship between a conventional diode and an amplifying diode. As shown in the left panel of Fig. 1(a), for the acoustic wave incident from the left-hand side (defined as the forward direction), the transmitted wave has a unit amplitude ($|t_f| = 1$), while the backward transmitted wave is completely suppressed ($t_b = 0$), exhibiting nonreciprocal wave propagation. The right panel of Fig. 1(a) shows the requirement of the proposed amplifying diode. For a forward incident wave, we have amplification in transmission and zero reflection simultaneously. It has to be achieved by some active Huygens (secondary) sources in the middle to generate scattering waves only in the forward direction, while adding in-phase to the incident wave ($|t_f| > 1$). The zero back-scattering generates no reflection ($r_f = 0$). For a backward incident wave, the same active Huygens sources still generate zero back-scattering ($r_b = 0$) while adding out-of-phase to the incident waves to suppress the backward transmission ideally to a value zero ($t_b = 0$). These Huygens sources have to be both active and asymmetric and are thus linked to the active Willis coupling in an acoustic effective medium setting for a one-dimensional (1D) metamaterial.

Without losing generality, the 1D metamaterial is along the $x$-direction and has a $2 \times 2$ constitutive matrix $\{\{\beta, i\tau\}, \{i\tau', \rho\}\}$. $\beta$ and $\rho$ represent the dimensionless compressibility and density in relative to air. $\tau$ and $\tau'$ are the required Willis coupling terms. Under an $e^{-i\omega t}$ time convention, the wave equation can be written as

$$\partial_x \begin{pmatrix} v \\ p \end{pmatrix} = \frac{i\omega}{c} \begin{pmatrix} \beta & i\tau \\ i\tau' & \rho \end{pmatrix} \begin{pmatrix} p \\ v \end{pmatrix}, \quad (1)$$

where $p$ is the pressure field, $v$ is the velocity field multiplied by the acoustic impedance of air and $c$ is the sound speed of air. To obtain an analytic insight, we first consider the scattering from only a small section of the metamaterial of thickness $L$ in air. In this case, the scattering matrix of the metamaterial can be obtained by integrating Eq. (1) with the Padé's approximation [28] to become

$$\begin{pmatrix} t_f & r_b \\ r_f & t_b \end{pmatrix} \cong \begin{pmatrix} 1 + \frac{i\phi_0}{2}(\beta + \rho + 2i\tau_{nr}) & \frac{i\phi_0}{2}(\beta - \rho - 2i\tau_r) \\ \frac{i\phi_0}{2}(\beta - \rho + 2i\tau_r) & 1 + \frac{i\phi_0}{2}(\beta + \rho - 2i\tau_{nr}) \end{pmatrix}, \quad (2)$$

where the subscript "f" ("b") denotes the transmission and reflection coefficients in the forward (backward) direction. $\phi_0 = \omega L/c$ is the phase elapse across air with the same thickness of the metamaterial (as one meta-atom later) and is much less than $2\pi$. We have also decomposed the Willis coupling terms into the reciprocal $\tau_r$ and non-reciprocal components $\tau_{nr}$, so that $\tau = \tau_r + \tau_{nr}$, and $\tau' = -\tau_r + \tau_{nr}$ [10]. According to Eq. (2), a non-zero $\tau_r$ allows asymmetric reflections in the forward and backward directions, while a nonzero $\tau_{\mathrm{nr}}$ offers us a tuning knob for non-reciprocal transmission with $t_f - t_b = -2\phi_0 \tau_{\mathrm{nr}}$. To avoid potential disturbance back to the source or other devices from the reflected signal, we choose to set $r_f = r_b = 0$ in Eq. (2), which gives rise to the impedance matching condition:

$$\tau_r = 0, \beta = \rho. \tag{3}$$

More specifically, a non-zero $\tau_r$ means a non-zero $r_f - r_b$ while $\beta \neq \rho$ means a non-zero $r_f + r_b$. An impedance-matched metamaterial has thus a purely non-reciprocal Willis coupling $\tau = \tau' = \tau_{\mathrm{nr}}$, with transmission coefficients

$$t_f \cong 1 + i\phi_0(i\tau_{\mathrm{nr}} + \beta),$$
$$t_b \cong 1 + i\phi_0(-i\tau_{\mathrm{nr}} + \beta).$$

Since there is no reflections at each small section, cascading same copies of the section (or meta-atoms) simply turns the scattering coefficients of the metamaterial into the following form:

$$t_f e^{\tau_{nr}\phi_0} = t_b e^{-\tau_{nr}\phi_0} = e^{i\beta\phi_0},$$
$$r_f = r_b = 0. \tag{4}$$

This gives further flexibility in controlling the amplification factor on the incident waves. By requesting a real value of $\tau_{nr}$, the waves are exponentially amplifying and decaying in opposite directions with zero reflection. By cascading more atoms, it will approach the required amplifying diode operation.

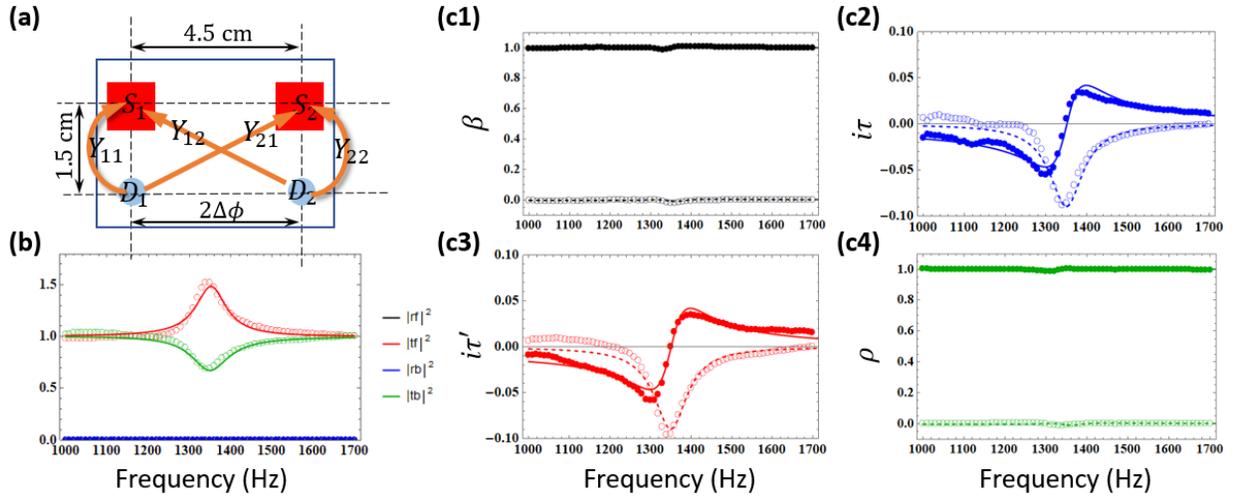

**Figure 2** (a) Schematic representation of the meta-atom consisting of two speakers (labeled as $S_1$ and $S_2$) and two microphones ($D_1$ and $D_2$). The phase distance between two speakers (microphones) is $2\Delta\phi$. Orange arrows connecting the microphone $D_j$ to a speaker $S_i$ represent the time-convolution labeled by $Y_{ij}$. (b) The transmittance and reflectance for both the forward and backward directions, validate the direction-dependent amplification phenomenon and zero reflection. The symbols (lines) show the experimental (analytical) results. (c1-c4) The extracted four constitutive parameters from the scattering parameters in (b), verifying the target constitutive parameters: $\tau = \tau' = \tau_{nr}$ and $\beta = \rho$. The solid (open) symbols represent the real (imaginary) part of the experimental results, and the solid (dashed) lines denote the real (imaginary) part of the analytical results.

To realize the amplifying diode, each meta-atom has to respond to both monopolar and dipolar incoming waves so that the monopolar and dipolar scattering waves behave as the described Huygens source. We adopt the programmable metamaterial approach shown in Fig. 1(b), with totally nine meta-atoms, with a period of $L = 9$ cm in a 1D waveguide. Each meta-atom consists of a pair of speakers and microphones interconnected by a microcontroller on an integrated unit (the two insets). It can at most carry out four different channels of time-convolution in connecting the two speakers (labeled as $S_1$ and $S_2$ in unit of pressure) to the two microphones (labeled as $D_1$ and $D_2$ in unit of pressure) as a matrix multiplication in the time-harmonic representation:

$$\begin{pmatrix} S_1 \\ S_2 \end{pmatrix} = \begin{pmatrix} Y_{11} & Y_{12} \\ Y_{21} & Y_{22} \end{pmatrix} \begin{pmatrix} D_1 \\ D_2 \end{pmatrix}. \tag{5}$$

Such a representation is also schematically shown in Fig. 2(a). The distance between the two speakers or the two microphones is 4.5 cm in the actual implementation, labeled as $2\Delta\phi$ in terms of phase elapse in air. In the following, we set $\beta = \rho = 1$ for simplification and $\tau_{nr} = -0.09$. It can be further proved that the required scattering coefficients (Eq. (4)) can be realized now by zero $Y_{11}$ and $Y_{22}$, together with

$$Y_{12} = ie^{\tau_{nr}\phi_0/2} \sinh(\tau_{nr}\phi_0/2) \csc 2\Delta\phi, \tag{6}$$
$$Y_{21} = -ie^{-\tau_{nr}\phi_0/2} \sinh(\tau_{nr}\phi_0/2) \csc 2\Delta\phi,$$

which are implemented within the microcontroller of each meta-atom through two time-domain convolutions or equivalently a Lorentzian-type resonance in the frequency domain [19,21]:

$$Y_{12/21}(f) = \frac{g_{12/21}f_0}{f_0^2 - (f + i\gamma)^2}. \tag{7}$$

Then, $Y_{12}$ and $Y_{21}$ can become almost purely imaginary if we choose the operation frequency at $f_0$, the resonating frequency at 1350 Hz in this work. We also set the resonance linewidth $\gamma = 50$ Hz. Finally, the resonance strengths are set by $g_{12} \cong 2i\gamma Y_{12}(f_0) = -10.1$ Hz and $g_{21} \cong$

$2i\gamma Y_{21}(f_0) = 12.3$ Hz. More details are given in Supplementary Material [29] for the implementation model.

The transmittance and reflectance for a single unit cell in both the forward and backward directions are measured and shown in Fig. 2(b) as symbols. They match well with the analytic model results (given by Eq. (7)) shown as solid lines. Around the resonance frequency, the forward transmittance (red color) of the meta-atom is around 1.5, demonstrating that the forward transmitted signal is amplified. However, the backward transmittance (green color) is suppressed to around 0.68, exhibiting obvious transmission contrast resulted from non-reciprocity. In fact, for such a meta-atom, the forward transmission amplitude times the backward transmission amplitude is approximately equal to one as expected in the model. In addition, the reflection coefficients for both the forward and backward directions (black and blue colors) are almost zero, validating our target design. To confirm the Willis coupling, we also extract the four constitutive parameters of the metamaterial from the complex transmission and reflection coefficients [29], as shown in Fig. 2(c). The real and imaginary parts of the experimental results are shown with solid and open symbols respectively, agreeing with the analytical results shown with solid and dashed lines. The two measured Willis couplings $\tau$ and $\tau'$ are almost the same, i.e. purely non-reciprocal Willis coupling. At the resonance frequency 1350 Hz, the Willis couplings terms $i\tau$ and $i\tau'$ are well approximated to be purely imaginary numbers $-0.09i$, resulting in the maximum contrast between the forward and backward transmission amplitude in the spectrum. In addition, the extracted constitutive parameters also satisfy the impedance matching condition: $\beta = \rho \cong 1$ with zero $\tau_r$, giving us zero reflections in both the forward and backward directions. We also note that the constitutive parameters extracted for a single unit cell (of thickness $L$) can also be used for the situation when there are multiple unit cells as there is negligible near-field coupling between neighboring unit cells in this work.

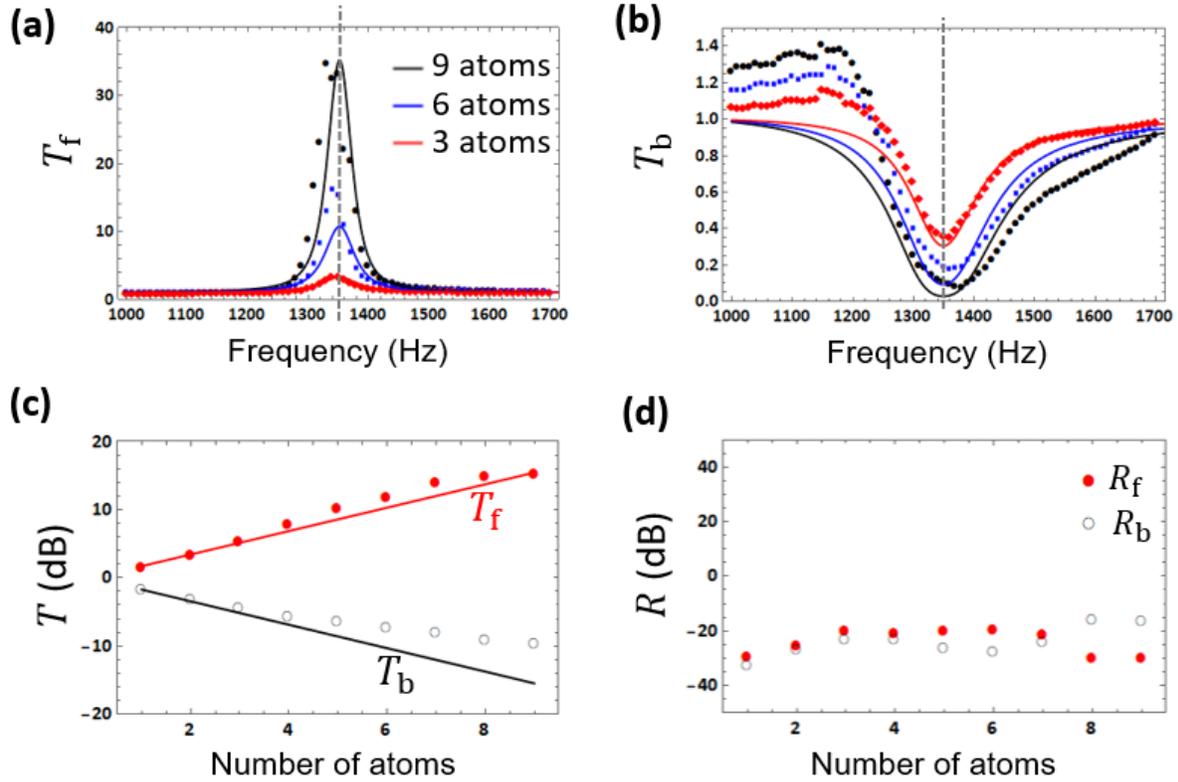

**Figure 3** The experimental results of (a) forward and (b) backward transmittance for the metamaterial amplifying diode consisting of 3 atoms (black symbols), 6 atoms (blue symbols) and 9 atoms (red symbols). (c) The forward (red color) and backward (black color) transmittance at the resonance frequency varies as the number of atoms increases from 1 to 9. (d) The measured reflectance at the resonance frequency for the forward and backward directions. The symbols (lines) represent the experimental (analytical) results.

Now, we cascade the same meta-atoms in the 1D waveguide to realize the amplifying diode. As shown in Fig. 1(b), 9 meta-atoms are equally spaced in the top cover of the 1D waveguide. In the experiment, we gradually increase the number of turned-on meta-atoms from 1 to 9, and then measure the scattering parameters for the 1D metamaterials consisting of the different number of meta-atoms. Figure 3(a) shows the forward transmittance of 1D metamaterials consisting of 3 meta-atoms (red color), 6 meta-atoms (blue color) and 9 meta-atoms (black color). Both the experimental (symbols) and analytical (lines) results show that, as the number of meta-atoms increases, the forward transmitted signal has a larger amplification ratio around the resonance frequency. For a metamaterial consisting of 9 atoms, the transmission intensity in the forward direction can go up to around 35. In contrast, in the backward direction, the backward transmitted signal has a lower intensity as the number of meta-atoms increases, and is almost completely suppressed for the metamaterial consisting of 9 atoms, i.e. nearly perfect absorption occurs for the backward incident wave. The huge contrast between the forward and backward transmission intensity demonstrates a significant isolation effect using such an active metamaterial with 9 atoms. To closely study the relationship between the transmission intensity and the number of meta-atoms, we focus on the transmittance at resonance frequency 1350 Hz (denoted by dashed lines in Figs. 3(a) and (b)). Figure 3(c) plots the transmission intensity (in $10 \log T$) at the resonance frequency for the forward (red color) and backward directions (black color) against the number of meta-atoms. The transmission intensity in the forward direction is exponentially amplified as the number of meta-atoms increases, while the transmission intensity in the backward direction is exponentially decayed. For the metamaterial with $\tau = \tau' = \tau_{nr}$ and $\beta = \rho \approx 1$, the exponentially amplification or decay of the transmittance depends on the Willis parameter $\tau$ as $\pm 8.686(-\phi_0\tau)n$ where $n$ represents the number of meta-atoms. On the other hand, the measured intensity of reflected signals in both the forward and backward directions is independent of the number of meta-atoms, and keeps at low intensity (around -20 dB) as the number of meta-atoms increases, as shown in Fig. 3(d). Therefore, based on the active meta-atoms, we experimentally realize the amplifying diode with a giant amplification ratio for the forward transmitted signals.

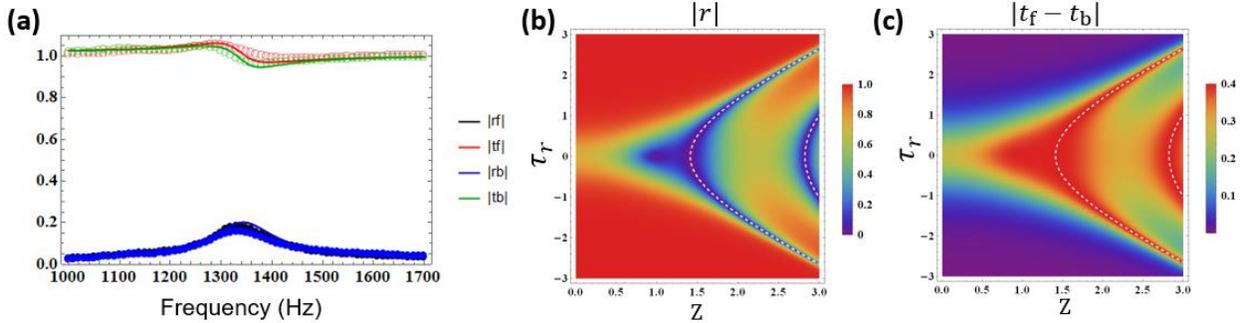

**Figure 4** (a) The measured amplitude of the scattering coefficients (symbols) for a meta-atom with $\tau = \tau'$ but $\beta \neq \rho$, agreeing to the analytical results (lines). (b) The reflection amplitude $|r|$ and (c) the transmission amplitude contrast $|t_f - t_b|$ for a general Willis metamaterial with thickness $L$ against the parameter $Z = \sqrt{\rho/\beta}$ and the reciprocal Willis parameters $\tau_r$. The white lines indicate the FP resonance condition and the point at $Z = 1$ and $\tau_r = 0$ indicates the amplifying diode working condition with impedance matching.

In the above implementation of the amplifying diode, we have only adopted a purely non-reciprocal Willis coupling $\tau_{nr}$ with impedance matching: $\beta = \rho$ and $\tau_r = 0$. We provide further analysis on the scattering properties for more general Willis metamaterials. As a simple example, we flip the sign of resonance strength of the kernel $Y_{12}$, i.e. $g_{12} = 10$ Hz. The small contrast between $Y_{12}$ and $Y_{21}$ results in a much smaller magnitude of $\tau_{nr}$ and thus a smaller transmission contrast between the forward and backward directions, as shown in Fig. 4(a). There are also some reflections with $r_f = r_b$ due to the deviation from the impedance matching condition as $\beta \neq \rho$ with $\tau_r = 0$. In this case, the reflection coefficient can be expressed as

$$r_\text{f} = r_\text{b} = \frac{Z^2 - 1}{2iZ \cot(n_r\phi_0) + 1 + Z^2}, \tag{8}$$

where $n_r = \sqrt{\rho\beta}$ and $Z = \sqrt{\rho/\beta}$ are the refractive index and the impedance of the medium, being independent of the purely non-reciprocal Willis coupling term $\tau_{nr}$. This is in contrast with the reciprocal Willis metamaterials where the Willis coupling affects the relative refractive index and split the impedance into two values depending on the direction.

Now we extend to a general Willis metamaterial with both the reciprocal and non-reciprocal components of the Willis coupling present. Here, we fix $\beta = 1$ and $\tau_{nr} = -0.09$ as before while we scan $Z$ (in fact $\sqrt{\rho}$) and $\tau_r$. In the current case with all real numbers for $\beta, \rho, \tau_r$ and $\tau_{nr}$, a non-zero $\tau_r$ only causes the two reflection coefficients in opposite directions differ in phase but not in magnitude, we plot the reflection amplitude $|r|$ and the amplitude of the transmission contrast $|t_\text{f} - t_\text{b}|$, against $Z$ and $\tau_r$ in Fig. 4(b) and (c). At $\tau_\text{r} = 0, Z = 1$, a clear dip in $|r|$ with a peak at $|t_\text{f} - t_\text{b}|$ indicates the demonstrated case of amplifying diode. In fact, the denominator in Eq. (8) suggests the possible occurrence of Fabry-Pérot (FP) resonance, which now becomes the discrete "parabolic" bands in getting $|r| = 0$, whose positions are denoted as the white dashed curves in both Fig. 4(b) and 4(c). It is further noted that the transmission contrast $|t_\text{f} - t_\text{b}|$ also has a large amplitude at FP resonances.

Unlike the impedance matching approach (the demonstrated case), the FP resonance approach will be useful in a situation that we do not have a full control on the Willis couplings so that both reciprocal and non-reciprocal Willis couplings are present at the same time. The FP approach can only work at a chosen thickness to satisfy resonance condition (in a way similar to the wavelength sensitivity for a FP resonance). On the contrary, the impedance matching approach allows amplification cascaded as thickness grows, i.e. with a tunability on the amplification ratio and a higher tolerance on the operation condition. It will be useful in a situation that the gain of individual metamaterial atom is limited by its power source. It is also interesting to note that the current implementation uses non-Hermitian components, which is also related to a concept called parity-time symmetric laser absorber [30-33]. An optical medium can act as a laser or a perfect absorber depending on different coherent wave excitations. In the current work, we have actually demonstrated a similar concept in acoustics that the Willis medium acts as an amplifier or an absorber depending on the way of excitation.

In summary, we have proposed and demonstrated the concept of an acoustic amplifying diode from either impedance matching or from FP resonance using non-reciprocal and active Willis coupling. An incident wave will be amplified without reflection in one direction while completely absorbed in another direction. The demonstrated diode action with zero reflection will be immediately useful for sound isolation. The additional amplifying effect in one direction helps to construct ultrasensitive sensors. Looking forward for applications in larger systems, the ability of distributing the amplification (without reflection) into different components and without using time-modulation will further enable non-reciprocal control of wave propagation in more general settings, such as control in two-dimensional setting for non-reciprocal communication and non-Hermitian topology [34,35].

**Acknowledgment**

The work is supported by Hong Kong RGC project no. 16303019 and AoE/P-502/20.